\documentclass[a4paper]{article}
\usepackage{graphicx}
\begin{document}
\title{Sum rules for total hadronic widths of mesons}
\author{Micha{\l} Majewski\\
Dept. of Theoretical Physics, Univ. of Lodz\\ Pomorska~149/153, 90-236~\L\'od\'z, Poland}
\date{}
\maketitle
\begin{abstract}
Mass sum rules for meson multiplets derived from exotic commutators may be written for complex masses. Then the real parts give the well known mass formulae (GM-O, Schwinger, Ideal) and the imaginary ones give the corresponding sum rules for total hadronic widths. The masses and widths of the meson nonets submit to a definite orders. It thus follows that tables of the meson nonets should include information about masses, widths and the orders as well as the mixing angle. The width sum rule for the nonet complying with Schwinger mass formula may be depicted as a straight line in the $(m,\Gamma)$ plane. It is easily verifiable and satisfied better for high mass nonets.
\end{abstract}

\section{Introduction}
The particle width is one of its main characteristics as much important as mass and 
discreet quantum numbers. It tells us something different than the mass and sometimes it 
may tell more: the widths of the particles with similar masses may differ by many orders; then the widths inform us first which interaction---strong, electromagnetic or weak is responsible for their decay. For hadronic decays the differences are not so big, but usually are of the same order as mass differences. Therefore they merit attention.

The difficulty with the widths within the meson multiplet is that they are in a sense accidental. Indeed, selection rules may suppress more or less the decay of a particular particle thus destroying any given regularity. Such an effect should be especially transparent in low mass multiplets where for some particle two-body decays are forbidden and many-body decays are suppressed (e.g. $\omega$-meson). For more massive multiplets, where many decay channels are opened, we may expect better agreement. However the prediction may be interesting in any case. 
       
\section{Sum rules for nonets}
\label{sec:1}
The approach is based on the technique of exotic commutators \cite{O}. 
The following system of mass sum rules for a nonet has been obtained \cite{1}:
\begin{eqnarray}
l_1^2 + l_2^2 &=& 1\label{1.1}\\
l_1^2 z_1 + l_2^2 z_2 &=& \frac{1}{3} a + \frac{2}{3} b\; (\equiv z_8)\label{1.2}\\
l_1^2 z_1^2 +l_2^2 z_2^2 &=&\frac{1}{3} a^2 +\frac{2}{3} b^2\label{1.3}\\
l_1^2 z_1^3 +l_2^2 z_2^3 &=&\frac{1}{3} a^3 +\frac{2}{3} b^3\label{1.4}
\end{eqnarray}
Here $a$, $K$, $z_1$, $z_2$ stand for the mass squared of the isotriplet, isodublet and 
isoscalar physical mesons respectively ($z_1<z_2$ by choice),  $z_8$ is the GM--O 
mass squared, $b=2 K - a$ and the real coefficients $l_1$, $l_2$ are introduced by the equation: $|z_8\rangle=l_1 |z_1\rangle +l_2 |z_2\rangle$. The known mass formulae for a nonet follow from eqs. (\ref{1.1})--(\ref{1.4}).

Eqs. (\ref{1.1})--(\ref{1.4}) may be considered for complex masses
$\hat{M}^2$=$\hat{m}^2-i\hat{m}\hat{\Gamma}$.
$\hat{M}^2$ is now nonhermitean, but it can be diagonalized and has orthogonal eigenfunctions. We use for the masses squared the notations: $a$, $K$, $x_1$, $x_2$, $x_8$, $b$ and for the appropriate imaginary parts the notations: $\alpha$, $\kappa$, $y_1$, $y_2$, $y_8$, $\beta$(=$2\kappa -\alpha $=${\sqrt(b)}\Gamma _b$). The coefficients $l_1$, $l_2$ are complex numbers, but $l^{2}_{1}$, $l^{2}_{2} $, which are their modula squared, are real. 

The real parts of the masses squared satisfy usual mass formula and the imaginary ones give the sum rules for the widths. 

For GM-O nonet (follows from eqs. (\ref{1.1}), (\ref{1.2})) we find
\begin{equation}
l_{1}^{2}=\frac{x_{2}-x_{8}}{x_{2}-x_{1}}, \quad
l_{2}^{2}=\frac{x_{8}-x_{1}}{x_{2}-x_{1}}, \quad
\frac {y_2-y_8}{x_2-x_8}=\frac {y_8-y_1}{x_8-x_1}.\label{2.3}
\end{equation}

For Schwinger nonet (follows from eqs. (\ref{1.1})--(\ref{1.3})) we have
\begin{equation}
\frac {y_2-y_8}{x_2-x_8}=\frac {y_8-y_1}{x_8-x_1}=\frac {\beta -\alpha }{b-a} \label{d.1}
\end{equation}
and separate equations for the real and imaginary parts of the mass operator:  
\begin{eqnarray}
(a-x_1)(a-x_2)+2(b-x_1)(b-x_2)&=&0\label{d.2}\\
(\alpha -y_1)(\alpha -y_2)+2(\beta -y_1)(\beta -y_2)&=&0.\label{d.3}
\end{eqnarray}
For these nonets only two mass orders are allowed and two widths orders for each of them: 
\begin{eqnarray}
a<x_1<b<x_2;\quad \alpha >y_1>\beta >y_2;\quad 
{\rm  or}\quad  \alpha <y_1<\beta <y_2\label{d.4}\\
x_1<a<x_2<b;\quad y_1>\alpha >y_2>\beta;\quad
{\rm or}\quad  y_1<\alpha <y_2<\beta .\label{d.5}
\end{eqnarray}
Both mass orders are observed \cite{2}. 

Some of the well established Schwinger nonets are collected in the Table 1.

For the Ideal nonet (folows from eqs. (\ref{1.1})-- (\ref{1.4})) we have
\begin{equation}
x_1=a,\quad x_2=b,\quad y_1=\alpha ,\quad y_2=\beta .\label{d.6}
\end{equation}

If we apply eqs. (\ref{1.1})--(\ref{1.3}) to the octet states ($l_{1}^{2}=1$, $l_{2}^{2}=0$) we get degenerate octet (all mases and widths identical). Degenerate multiplets (octets(?), nonets(?)) do exist. They are shown in the Table 2 .

The formula (\ref{d.1}) shows that the points $(m^2,m\Gamma )$ of the
Schwinger nonet states lie on a straight line in the $(m^2,m\Gamma)$ plane
and consequently the poins $(m,\Gamma)$ lie on the straight line in the plane
$(m,\Gamma)$. The slope of this line is indefinite. Mass-width diagram for the 
nonet $2^{++}$ is shown on the Figure 1. For the nonet $1^{- -}$ the agreement is worse, 
for the nonet $3^{- -}$ it is quite good.

\begin{table}
\caption{Some well established nonets of mesons (masses and widths in MeV)}
\label{tab1}
\footnotesize
\addtolength{\topmargin}{-25pt}
\addtolength{\textheight}{25pt}
\begin{tabular}{|r@{}l||c||c|c||c|c|}
\hline
&&$m_K$&$m_a$&$m_1$&$m_b$&$m_2$\\
\multicolumn{2}{|c||}{$J^{PC}$}&$\Gamma_K$&$\Gamma_a$&$\Gamma_1$
&$(2\kappa -\alpha)m^{-1}$
&$\Gamma_2$\\
\hline
\multicolumn{2}{|c||}{multiplet}&\multicolumn{1}{c}{$\theta^{GMO}$}&
\multicolumn{2}{c}{mass order}&\multicolumn{2}{c|}{width order}\\
\hline\hline
\multicolumn{2}{|c||}{$1^{--}$}&$893.88\pm 0.26$&$769.3\pm 0.8$&$782.57\pm 0.12$&
$1.0031\pm 0.0011$&$1019.417\pm 0.014$\\
\cline{1-2}
$\bullet$&$\rho(770)$&&&&&\\
$\bullet$&$K^*(892)$&$50.7\pm 0.8$&$150.2\pm 0.8$&$8.44\pm 0.09$&
$-24.8\pm 2.1$&$4.458\pm 0.032$\\
$\bullet$&$\omega(782)$&&&&&\\
\cline{3-7}
$\bullet$&$\Phi(1020)$&\multicolumn{1}{c}{$(39.28\pm 0.16)^\circ$}&
\multicolumn{2}{c}{$a<x_1<b<x_2$}&
\multicolumn{2}{c|}{$\alpha>y_1>\beta>y_2$}\\
\hline\hline
\multicolumn{2}{|c||}{$2^{++}$}&$1429.0\pm 1.4$&$1318.0\pm 0.6$&$1275.4\pm 1.2$&
$1532.0\pm 3.1$&$1525\pm 5$\\
\cline{1-2}
$\bullet$&$a_2(1320)$&&&&&\\
$\bullet$&$K_2^*(1430)$&$103.8\pm 4.0$&$107\pm 5$&$185.1^{+3.4}_{-2.6}$&
$101.5\pm 11.9$&$76\pm 10$\\
$\bullet$&$f_2(1270)$&&&&&\\
\cline{3-7}
$\bullet$&$f'_2(1525)$&\multicolumn{1}{c}{$(30.67^{+1.56}_{-1.72})^\circ$}&
\multicolumn{2}{c}{$x_1<a<x_2<b$}&
\multicolumn{2}{c|}{$y_1>\alpha>y_2>\beta$}\\
\hline\hline
\multicolumn{2}{|c||}{$3^{--}$}&$1776\pm 7$&$1691\pm 5$&$1667\pm 4$&
$1857\pm 11$&$1854\pm 7$\\
\cline{1-2}
$\bullet$&$\rho_3(1690)$&&&&&\\
$\bullet$&$K_3^*(1780)$&$159\pm 21$&$161\pm 10$&$168\pm 10$&
$158\pm 53$&$87^{+28}_{-23}$\\
$\bullet$&$\omega_3(1670)$&&&&&\\
\cline{3-7}
$\bullet$&$\Phi_3(1850)$&\multicolumn{1}{c}{$(32.0^{+3,3}_{-7.5})^\circ$}&
\multicolumn{2}{c}{$x_1<a<x_2<b$}&
\multicolumn{2}{c|}{$y_1>\alpha>y_2>\beta$}\\
\hline\hline
\multicolumn{2}{|c||}{$1^{++}$}&$1341\pm 6$&$1230\pm 40$&$1281.9\pm 0.6$&
$1420\pm 0.012$&$1426.3\pm 1.1$\\
\cline{1-2}
$\bullet$&$a_1(1260)$&&&&&\\
$\bullet$&$K_A$&$134\pm 16$&$250\div 600$&$24.0\pm 1.2$&
$-447\div 89$&$55.5\pm 2.9$\\
$\bullet$&$f_1(1285)$&&&&&\\
\cline{3-7}
$\bullet$&$f_1(1420)$&\multicolumn{1}{c}{$35.26^\circ \div 41.00^\circ$}&
\multicolumn{2}{c}{$a<x_1<b<x_2$}&
\multicolumn{2}{c|}{$\alpha>y_1>\beta>y_2$}\\
\hline\hline
\multicolumn{2}{|c||}{$1^{+-}$}&$1322\pm 6$&$1229.5\pm 3.2$&$1170\pm 20$&
$1414\pm 9$&$1386\pm 19$\\
\cline{1-2}
$\bullet$&$b_1(1235)$&&&&&\\
$\bullet$&$K_B$&$135\pm 17$&$142\pm 9$&$360\pm 40$&
$130\pm 40$&$91\pm30$\\
$\bullet$&$h_1(1170)$&&&&&\\
\cline{3-7}
$\bullet$&$h_1(1380)$&\multicolumn{1}{c}{$0\div 35.26^\circ$}&
\multicolumn{2}{c}{$x_1<a<x_2<b$}&
\multicolumn{2}{c|}{$y_1>\alpha>y_2>\beta$}\\
\hline\hline
\end{tabular}
\end{table}

\begin{table}
\caption{Degenerate multiplets (masses and widths in MeV)}
\label{table2}
\small
\begin{tabular}{|r@{}l||c|c|c|}
\hline
&&$m_a$&$m_K$&$m_1$\\
\multicolumn{2}{|c||}{$J^{PC}$}&$\Gamma_a$&$\Gamma_K$&$\Gamma_1$\\
\hline
&$4^{++}$&&&\\
$\bullet$&$a_4(2040)$&$2014\pm 15$&$2045\pm 9$&$2034\pm 11$\\
$\bullet$&$K_4(2045)$&&&\\
$\bullet$&$f_4(2050)$&$361\pm 50$&$198\pm 30$&$222\pm19$\\
\hline
&$1^{--}$&&&\\
$\bullet$&$\rho(1450)$&$1465\pm 25$&$1414\pm 15$&$1419\pm 31$\\
$\bullet$&$K^*(1410)$&&&\\
$\bullet$&$\omega(1420)$&$310\pm 60$&$232\pm 21$&$174\pm 60$\\
\hline
&$1^{--}$&&&\\
$\bullet$&$\rho(1700)$&$1700\pm 20$&$1717\pm 27$&$1649\pm 24$, \quad$1680\pm 20$\\
$\bullet$&$K^*(1680)$&&&\\
$\bullet$&$\omega(1650)$&$240\pm 60$&$322\pm 110$&$220\pm 35$, \quad $150\pm50$\\
$\bullet$&$\Phi(1680)$&&&\\
\hline
\end{tabular}
\end{table}

\begin{figure}
\includegraphics[height=.45\textheight]{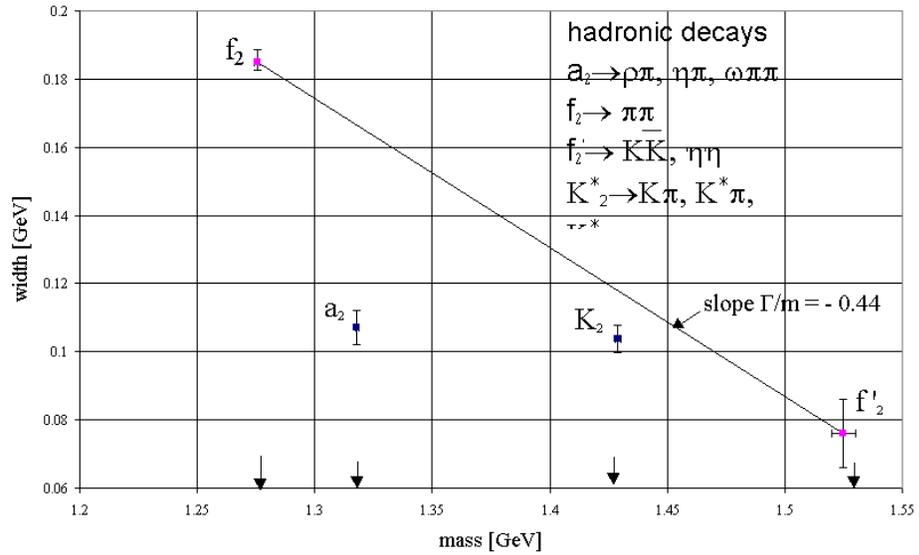}
\caption{Mass-width diagram of $2^{++}$ mesons---mass order:
$x_1 < a < x_2 <b$}
\end{figure}


\section*{Acknowledgments}
Valuable discusions with Profs. S.B. Gerasimov, P. Kosi\'nski and
V.A. Meshcheryakov are kindly acknowledged.

\end{document}